\documentclass{ifacconf}
\usepackage{mathrsfs}
\usepackage{graphicx}      % include this line if your document contains figures
\usepackage{natbib}        % required for bibliography
\usepackage{amsmath}
\usepackage{multirow}
\usepackage{amsfonts}
\usepackage{float}
\usepackage{amssymb}
\usepackage{algorithm,float}
\usepackage{algpseudocode}

\newtheorem{theorem}{Theorem}
\newtheorem{lemma}{Lemma}

\newtheorem{definition}{Definition}
\newtheorem{remark}{Remark}
\begin{document}
\begin{frontmatter}

\title{Model Predictive Control for Neuromimetic Quantized Systems\thanksref{footnote}} 

\thanks[footnote]{Support from various sources including the Office of Naval Research grant number N00014-19-1-2571 is gratefully acknowledged.}

\author[First]{Zexin Sun} 
\author[First]{John Baillieul} 

\address[First]{Boston University, Boston, MA, 02215 USA (e-mail: $\{zxsun, johnb\}@bu.edu$).}

\begin{abstract}                % Abstract of not more than 250 words.
Based on our recent research on neural heuristic quantization systems, we propose an emulation problem consistent with the neuromimetic paradigm. This optimal quantization problem can be solved with model predictive control (MPC) by deriving the conditions under which the quantized system can guarantee (asymptotic) stability during emulation by optimizing a Lyapunov-like objective function. The neuromimetic model features large numbers of discrete inputs, and the optimization involves integer variables. The approach in the paper begins by solving an
optimization using model predictive control (MPC) and then using a neural network to train the data generated in this process and applying Fincke and Pohst's sphere decoding algorithm to narrow down the search for the optimal solution.
\end{abstract}

\begin{keyword}
Model predictive control, Quantized system, Emulation
\end{keyword}

\end{frontmatter}
%===============================================================================

\section{Introduction}
Model predictive control (MPC) is a powerful control technique in dynamic systems, power inverters, and dynamic reference trajectories, as shown in \cite{berberich2022linear}. It can predict future system behavior from the current system state by solving an optimal control problem at each sampling instant. As discussed in \cite{baillieul2019perceptual}, the evolution of a control theory for systems that exhibit the kind of resilience seen in neurobiology involves input and output signals generated by the collective activity of vast numbers of simple elements. This is one of the primary motivations for studying what we have called overcomplete control systems. Our goal is to understand systems whose overall functionality depends on discrete sets of inputs that operate collectively in groups. One such feature is control modulation involving actions of very large numbers of simple inputs and outputs that are effective in influencing the system dynamics only in their aggregate operation. Quantized input systems have been studied for many decades, such as in \cite{lewis1963optimum} and \cite{liu1990optimal}, but this work was primarily concerned with the digital round-off errors. The work reported in this paper is focused on developing quantization methods merged with MPC using simple inputs inspired by neurobiology to realize emulation.

Consider the linear time-invariant (LTI) systems of the form 
\begin{equation}
		\dot x(t)=Ax(t) + Bu(t), \ \ \ x\in\mathbb{R}^n, \ \ u\in\mathbb{R}^m,	
\end{equation} 
and use simple stablized feedback control law $u=Kx$, we obtain the closed-loop LTI system
\begin{equation}
	\dot x(t)=Hx(t),\ \ x(0)=x_0,\label{equation:LTI}
\end{equation} 
where $H=A+BK\in\mathbb{R}^{n\times n}$ is Hurwitz. Following \cite{baillieul2021neuromimetic}, we consider the problem of emulating (\ref{equation:LTI}) by a discrete-time system with quantized inputs
\begin{equation}
	x_q(k+1)=A_qx_q(k)+hB_qu(k),
	\label{equation:qs}
\end{equation}
where $x_q\in\mathbb{R}^n$ is the system state, $u\in\mathbb{U}=\{-1,0,1\}^m$ is the set of possible quantized inputs and $m\gg n$. $h$ is the time step.

In \cite{baillieul2021neuromimetic}, we formulated two types of emulation problems and only focused on the restricted problem. The general one is further discussed in this paper. The goal is to find piecewise constant quantized inputs with $h>0$ such that the resulting trajectories of (\ref{equation:qs}) with initial state $x_q(0)=x_0$ approximate the continuous system $(\ref{equation:LTI})$. Here, we solve the emulation problem that finds a partition of the state space $\{U_i\,:\, \cup\  U_i = \mathbb{R}^n;\ \ U_i^o\cap U_j^o=\emptyset;\ U_i^o={\rm interior}\ U_i\}$ and a selection rule depending on the current state $x_q(k)$ for assigning values  of the input at the $k$-th time step to be $u(k)\in{\cal U}=\{-1,0,1\}^m$, so that for each $x\in U_i$, the quantized system (\ref{equation:qs}) is as close as possible to the LTI system (\ref{equation:LTI}) according to an appropriate metric like the magnitudes (i.e., $< B_qu,Hx>$ if $A_q$ is identity) and directions (i.e., $||Hhx_0||-||hB_qu||$) as studied in \cite{sun2022neuromimetic}. 

In the present paper, we show how a Lyapunov-like objective function can be used to formulate a quantized system model predictive control (QS-MPC) theory along the line pursued in \cite{xu2022steady}. The problem of using a quantized system to optimally emulate systems that are continuous in both time and state variable is considered. Sufficient conditions guaranteeing the asymptotic stability of solutions to the optimal emulation problem are established. The computational complexity of the integer optimization problem is addressed by generating a neural network model and an appropriate least squares reformulation.

\section{MPC for Quantized Systems}
We know that the existence of a control Lyapunov function provides sufficient conditions for the existence of a controller that ensures asymptotical stability for a discrete-time system. Therefore, establishing the MPC's stability can be approached by finding a candidate-Lyapunov function as its cost function. In general, the cost function contains the terminal cost and stage cost with the form  \[J(x,U)=p(x_N)+\sum_{i=0}^{N-1}q(x_i,u_i),\] where we have a prediction horizon $N$ and the predicted input sequence $U=\{u_0,\dots,u_{N-1}\}$. The commonly used cost functions $p(\cdot)$ and $q(\cdot)$ are quadratic in the states and control inputs. Then, the QS-MPC approach can be formulated as solving the following optimization problem: 
\begin{equation}
	\label{MPC_optimal}
	\begin{split}
		\min\limits_{u_{0|k},...,u_{N-1|k} }&\quad J\left(x_q(k),U(k)\right)=|x_{N|k} -x_{ref}(N|k)|^2_P\\
		&\qquad\ \ \ \ \ \ +\sum_{n=0}^{N-1}|x_{n|k}-x_{ref}(n|k)|^2_Q+	 |u_{n|k}|^2_R	\\
		s.t. &\quad u_{n|k}\in\left\{-1,\ 0,\ 1\right\}^m,\\
		&\quad x_q(k)= x_{0|k},\\
		&\quad x_{n+1|k}=A_qx_{n|k}+hB_qu_{n|k},\\
		&\quad x_{ref}(n+1|k)=e^{Hh}x_{ref}(n|k),\\
		&\quad \quad \quad \quad \quad \quad \quad \quad \quad \quad \quad 	\forall n= {0,1,...,N-1},
	\end{split}
\end{equation}
where $P,Q, R$ are positive definite matrices, $h$ is the sampling interval and the function $|x|^2_P=x^TPx$, with similar definitions for $Q$ and $R$. The reference system $x_{ref}$ is the linear time-invariant system (\ref{equation:LTI}). Matrix $H=A+BK$ is the state matrix of $x_{ref}$ and could be thought of as specifying a target behavior of (\ref{equation:LTI}) to ensure the feedback law $u=Kx+v$ asymptotically steers the closed loop system toward the goal point $x_g$. Following general MPC procedures, the optimal input at time $k$ for the quantized system is $u^*(k)=u^*_{0|k}$, which is the first element in the predicted optimal input sequence $U^*(k)$.

As we generally establish the stabilizing MPC with a finite control set, we explore the convergence performance of our QS-MPC when conducting the emulation task. The following theorem gives a sufficient condition for stability.
\begin{theorem}
	\label{MPC:thm1}
	%Consider emulating a stable LTI system (\ref{equation:LTI}) utilizing the QS-MPC system (\ref{equation:qs}). If the optimization problem (\ref{MPC_optimal}) is always feasible, then the asymptotically stable solution for QS-MPC is guaranteed when the following conditions are satisfied:\\
	Consider emulating a stable LTI system (\ref{equation:LTI}) utilizing the QS-MPC system (\ref{equation:qs}). The asymptotically stable solution for QS-MPC is guaranteed when the following conditions are satisfied:\\
	(a) $A_q=e^{Hh}$;\\
	(b) the symmetric positive definite matrices $P, Q$ satisfy $Q-P+A_q^TP A_q\prec0$.
\end{theorem}
The proof of Theorem \ref{MPC:thm1} is given in the appendix.$\hfill\blacksquare$

\begin{remark}
	\nonumber
	Most quantized systems will not asymptotically converge to an equilibrium but only to a neighborhood of it. However, in this case, once the states approach the origin, the optimal input sequence can be zero, and the state transition matrix $A_q$ has all its eigenvalues inside the unit circle. Therefore, the QS-MPC is asymptotically stable.
\end{remark}
\begin{remark}
From Theorem \ref{MPC:thm1}, we find there are strict conditions on matrices $A_q,$ $P$, and $Q$. However, it is guaranteed that such $P, Q$ exist since $-P+A_q^TPA_q\prec0$. As long as the least eigenvalue of $P$ is much larger than the largest one of $Q$, we may find such matrices. 
\end{remark}
Next, we provide a relaxed condition to achieve this emulation but without the asymptotic tracking feature. Instead of setting $A_q=e^{Hh}$, we assume it is Schur stable. The main idea is to cancel the influence of $A_q$ matrix with $B_q$ in (\ref{equation:qs}). Since $B_q$ has a large number of columns, directions formed by $B_qu$ can have the dominant influence on $x_q(k+1)$. Without loss of generality, we assume the sampling interval is $h=1$.%Therefore, two conditions need to be satisfied: 1. all the states of the system from the initial point in the tracking process should be constrained in a set; 2. the quantized system can be closer to the origin, which means there exists a direction towards it, but it may not be the optimal direction solved by QS-MPC optimization problem.
\begin{lemma}
	\label{tracking:lemma1}
	When utilizing the quantized system $(\ref{equation:qs})$ to emulate the LTI system and without loss of generality, assume the equilibrium of the LTI is the origin. If $A_q$ is Schur stable, $(\ref{equation:qs})$ can track the stable LTI system (\ref{equation:LTI}) by solving the optimization problem (\ref{MPC_optimal}).
\end{lemma}
\begin{pf}
	Since $A_q$ is Schur stable, when there is no input (i.e., $u(t)=\mathbf{0}^m$), the quantized system will converge to the origin. By solving the optimization MPC problem (\ref{MPC_optimal}), it can decrease the error $\epsilon$ between these two systems, and the maximum error bound can be expressed as
	\begin{equation}
		\begin{split} 
			\epsilon_{max}\le&\max\{\max\{||A_q^tx_q(k)-e^{Ht}x_{ref}(k)||\}\},\\
			&\quad\quad\quad\quad \ \ \ \ t\in\{0,1,\dots,N-1\},k\in\mathbb{N}
		\end{split}
	\end{equation}
	where $\max\{||A_q^tx_q(k)-e^{Ht}x_{ref}(k)||\}$ represents the maximum error between two systems from time $k$ to the following predicted time $k+N$. %Since the trajectories of two systems are continuous, and error value $\epsilon$ between two systems is also continuous.
	 Because $A_q$ is Schur stable and $H$ is Hurwitz, when $k\to\infty$, $A_q^tx_q(k)\to 0$ and $e^{Ht}x_{ref}(k)\to0$ which implies $\epsilon\to0$. In addition, since $A_q^tx_q(k)$ and $e^{Ht}x_{ref}(k)$ are bounded, it is easily obtained that $||A_q^tx_q(k)-e^{Ht}x_{ref}(k)||$ is also bounded. Therefore, we can conclude there exists a constant can be the upper bound of $\epsilon_{max}$.$\hfill\blacksquare$
\end{pf}
\begin{definition}
	An open ball with center $p$ and radius $r>0$ can be written as $B_r(p)=\{x\in X:d(x,p)<r\}$.
\end{definition}
\begin{lemma}
	\label{tracking:lemma2}
	If there always exists a quantized input $\hat u$ to ensure $||A_qx_q(k)+hB_q\hat u||-||x_q(k)||\le0,\forall k$, the quantized system $(\ref{equation:qs})$ can track the stable LTI system (\ref{equation:LTI}) by solving the optimization problem (\ref{MPC_optimal}) and eventually converge to the set $\mathcal{X}=B_r(0)$.
\end{lemma}
Since if there exists a direction $\hat u(k)$ guiding the system to the origin at any time $k$, even though the quantized system selects other $u^*(k)$ to minimize the objective function, it can finally reach the origin as the LTI  system (\ref{equation:LTI}) approaches $\mathbf{0}$. The rigorous proof is omitted here.

From the above discussion, though solving problem (\ref{MPC_optimal}) can have satisfactory emulating performance as shown in the simulation in Section \ref{simulation}, it is a quadratic integer programming, which is NP-hard. Therefore, obtaining the optimal input sequence $U^*(k)$ is time-consuming, especially when the receding horizon $N$ becomes larger. In the following sections, we will show how to avoid this difficulty at the cost of some loss of accuracy.

\section{Reduced Complexity Suboptimal Solutions}
As stated in previous sections, solving integer programming is a computational challenge, and the directly rounding relaxed solutions of (\ref{MPC_optimal}) may lead to suboptimal solutions, which can negatively affect the emulating performance. In this section, we reformulate the problem to adapt the sphere decoding algorithm introduced in \cite{fincke1985improved} and utilized in \cite{hassibi2005sphere} and \cite{geyer2014multistep} based on the branch-and-bound method. Compared with the exhaustive enumeration method, the sphere decoding algorithm shrinks the size of the candidate control sequences by pruning the branch to improve efficiency. 
\subsection{Integer Least Squares Problem Formulation}
To apply the sphere decoding algorithm, the problem $(\ref{MPC_optimal})$ needs to be rewritten in the extensive form and constructed as an integer least squares problem. Define $ X_k=\begin{pmatrix}
	x_{0|k}\\
	x_{1|k}\\
	\vdots \\
	x_{N|k}
\end{pmatrix}\in\mathbb{R}^{(N+1)n}$, $U_k=\begin{pmatrix}
	u_{0|k}\\
	u_{1|k}\\
	\vdots \\
	u_{N-1|k}
\end{pmatrix}\in\{-1,0,1\}^{(N-1)m}$ and $R_k=\begin{pmatrix}
	x_{ref}(0,k)\\
	x_{ref}(1,k)\\
	\vdots \\
	x_{ref}(N,k)
\end{pmatrix}\in\mathbb{R}^{(N+1)n}$. Let $\tilde{A}=\begin{pmatrix}
	I\\
	A_q\\
	\vdots \\
	A_q^N
\end{pmatrix}$, $\tilde{B}=\begin{pmatrix}
	0& 0 & \cdots &0 \\
	B_q&  0& \cdots & 0\\
	A_qB_q&  B_q&  \cdots& 0\\
	\vdots&\vdots  & \ddots & \vdots\\
	A_q^{N-1}B_q& A_q^{N-2}B_q&\cdots  &B_q
\end{pmatrix}$, $\tilde Q=\begin{pmatrix}
	Q& 0 & \cdots &0 \\
	0&  Q& \cdots &0 \\
	\vdots& \vdots & \ddots &0 \\
	0& 0 & 0 &P
\end{pmatrix} $ and $\tilde R=\begin{pmatrix}
	R& 0 & \cdots &0 \\
	0&  R& \cdots &0 \\
	\vdots& \vdots & \ddots &0 \\
	0& 0 & 0 &R
\end{pmatrix}$. Then, $\tilde H=\tilde B^T\tilde Q\tilde B+\tilde R$ is positive definite and symmetric. The reconstructed problem $(\ref{MPC_optimal})$ is  
\begin{equation}
	\label{integer_least_H}
	\begin{split}
		\min\limits_{U_k}\quad& ||U_k+\tilde H^{-1}(\tilde Ax_q(k)-R_k)||_{\tilde H}^2\\
		s.t. \quad &u_{n|k}\in\left\{-1,\ 0,\ 1\right\}^m,\ \ \ \ \ \ \forall n= {0,1,...,N-1}. 
	\end{split}
\end{equation}
Detailed transformation steps can be found in the appendix. It is observed that the closed form solution for the unconstrained problem $(\ref{integer_least_H})$ is $U_{uncon K}=-\tilde H^{-1}$ $(\tilde Ax_q(k)-R_k)$. Since matrix $\tilde H$ is symmetric positive definite, there exists an invertible and lower triangular matrix $W\in \mathbb{R}^{Nm\times Nm}$ by Cholesky decomposition to factor $\tilde H=W^TW$ and $\tilde H^{-1}=W^{-1}W^{-T}$. Denote $\bar U_{uncon K}=WU_{uncon K}$. Then the optimization problem $(\ref{integer_least_H})$ has an integer least square objective function
\begin{equation}
	\begin{split}
	J\quad&=\quad(WU_k-\bar U_{uncon K})^T(WU_k-\bar U_{uncon K})\\
	&=\quad ||WU_k-\bar U_{uncon K}||_2^2.
	\end{split}
\end{equation}

\begin{figure}[h]
	\begin{center}
		\includegraphics[scale=0.55]{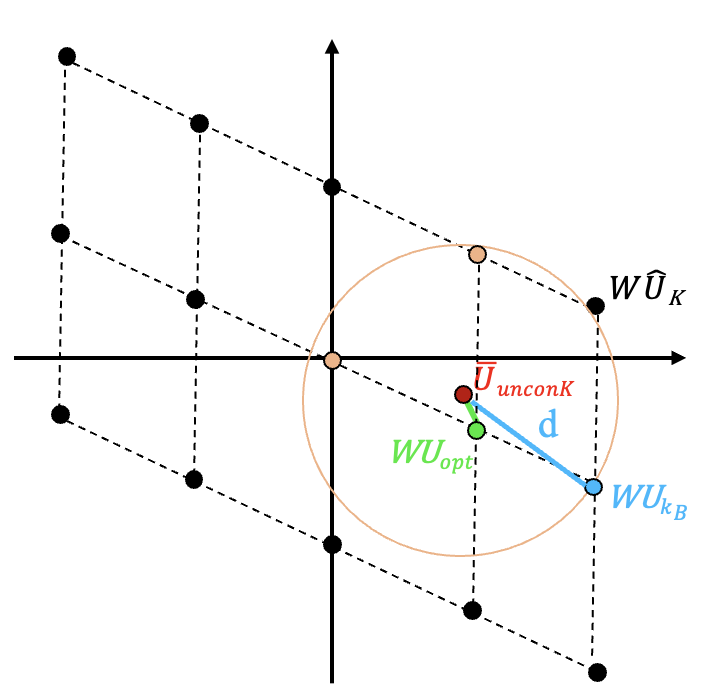}
	\end{center}
	\caption{ A diagram of sphere decoding algorithm. The red point denotes the unconstraint optimal solution and the green one is the optimal integer solution.}
	\label{fig:sphere}
\end{figure}

The various modified sphere decoding algorithms can be applied to solving this problem as was done in \cite{karamanakos2015computationally}. Though the sphere decoding algorithm can find the optimal solution by traversing a tree instead of applying an exhaustive search, the complexity of this algorithm depends on the radius $d$ of the sphere. Here, we choose $d=\min\{||WU_{k_B}-\bar U_{uncon K}||_2^2,||W\hat U_{k}-\bar U_{uncon K}||_2^2 \}$ in the initial, where $U_{k_B}$ is the direct integer lattice round-off of $U_{uncon K}$ and $\hat U_{k}$ is a shifted input sequence introduced in the next section. It guarantees that the radius is as small as possible and that there is at least one lattice point on or inside the sphere. Fig. \ref{fig:sphere} gives a schematic diagram.  It illustrates that instead of enumerating all control sequences, only points located in the orange circle (sphere) with center $\bar U_{unconK}$ and radius of $d$ will be considered, which shrinks the size of candidates to reduce the running time of (\ref{MPC_optimal}).

\subsection{Learning Optimal Activation Patterns Using Neural Networks}
\label{sub:machine learning}
While using the sphere decoding algorithm, we also collect data according to possible approximate metrics as in \cite{baillieul2021neuromimetic}, where we compared vectors in terms of both magnitude and direction. In terms of this metric, the difference between the quantized system state (\ref{equation:qs}), as determined by the MPC optimal input $u^*_{0|k}$ and the state of the LTI system can be recorded at each step. Applying our MPC approach to optimal emulation from different initial points provides a large amount of data, which we then use to train a neural network.  The trained model can choose the activation pattern directly at each step based on the current metric value of the two systems without solving the MPC problem. As illustrated in Section \ref{simulation}, the model based on classification is trained in a simple way but still can efficiently solve the emulation problem.

\section{Suboptimal QS-MPC Algorithm}
In this section, we propose an algorithm that can compute the suboptimal input sequence for each iteration instead of solving the original quadratic integer programming by relaxing the constraints. The relaxed quadratic programming problem is
\begin{equation}
	\label{relax MPC_optimal}
	\begin{split}
		\min\limits_{u_{0|k},...,u_{N-1|k} }&\quad J\left(x_q(k),U(k)\right)=|x_{N|k} -x_{ref}(N,k)|^2_P\\
		&\qquad\qquad+\sum_{n=0}^{N-1}|x_{n|k}-x_{ref}(n,k)|^2_Q+	 |u_{n|k}|^2_R	\\
		s.t. &\quad u_{n|k}\in[-1,1]^m,\\
		&\quad x_q(k)=x_{0|k},\\
		&\quad x_{n+1|k}=A_qx_{n|k}+B_qu_{n|k},\\
		&\quad x_{ref}(n+1,k)=e^{Hh}x_{ref}(n,k),\\
		&\quad \quad \quad \quad \quad \quad \quad \quad \quad \quad \quad 	\forall n= {0,1,...,N-1}.
	\end{split}
\end{equation}
After solving this problem and following \cite{grotschel2012geometric}, Babai estimation is used to round the solution to the nearest integer input sequence $\tilde U$ of vectors from the set $\{-1,0,1\}^m$. Then the algorithm estimates the suboptimal solution $\tilde U^*$ by choosing either the shifted optimal sequence $\hat U$ or the round-off sequence $\tilde U$. \\
\begin{algorithm}
	\caption{Suboptimal QS-MPC algorithm for emulating LTI systems}
	\label{Suboptimal QS-MPC}
	\begin{algorithmic}[1]
		\State Apply MPC with horizon predicted window size $N$ to solve problem (\ref{relax MPC_optimal}) and compute the optimal input sequence $U^*(0)= \{u^*_{0|0},...,u^*_{N-1|0}\}$; 
		\State Estimate $\tilde u^*_{i|0},\forall i\in\{0,1,...,N-1\}$ to be $\{-1,0,1\}^m$ by Babai estimation method. The suboptimal solution is denoted by  $\tilde U^*(0)=\{\tilde u^*_{0|0},...,\tilde u^*_{N-1|0}\}$ and take the first element $\tilde u^*_{0|0}$ as the input to the quantized system;
		\State Initialize the current iteration k=0, and input the total number of iterations $K$;
		\While{$k < K$}
		\State Define the shifted sequence $\hat U(k+1)=\{\tilde u^*_{1|k},...,\tilde u^*_{N-1|k},0\}$; 
		\State Solve the problem (\ref{relax MPC_optimal}) using MPC to compute the optimal input sequence $U^*(k+1)= \{u^*_{0|k+1},...,u^*_{N-1|k+1}\}$, and estimate it to be $\tilde U(k+1)=\{\tilde u_{0|k+1},...,\tilde u_{N-1|k+1}\}$;
		\If {$J(x(k+1),\hat U(k+1))\le J(x(k+1),\tilde U(k+1))$}			
		\State	$\tilde U^*(k+1)=\hat U(k+1)$;
		\Else
		\State $\tilde U^*(k+1)=\tilde U(k+1)$;
		\EndIf	
		\State Take the first element $\tilde u^*_{0|k+1}$ to the quantized system;
		\State $k\gets  k+1$;
		\EndWhile
	\end{algorithmic}
	
\end{algorithm}

\begin{theorem}
	\label{MPC:thm2}
	When conditions in Theorem \ref{MPC:thm1} are satisfied, Algorithm \ref{Suboptimal QS-MPC} solves  the optimal emulation problem for any stable LTI continuous time system (\ref{equation:LTI}). The optimal quantized emulation asymptotically approaches the LTI system equilibrium.
\end{theorem}

\begin{pf}
	To prove the convergence of this algorithm is the same as proving the cost function $J$ is a candidate Lyapunov function, which means $J(x_q(k),\tilde U^*(k))$ is strictly decreasing until it becomes 0. From the proof of Theorem \ref{MPC:thm1}, we obtain that when conditions (a) and (b) are satisfied, $J(x_q(k+1),\tilde U(k+1))\le J(x_q(k),\tilde U^*(k))$. After sufficiently many iterations, there is a $k$ such that this inequality become an equality, after which $ \forall l\ge 0,J(x_q(k+l),\tilde U(k+l))=0$. Therefore, similarly, from time $k$, the optimal input sequence of optimization problem (\ref{relax MPC_optimal}) is $\bold{0}$, and at this time $J(x_q(k+1),\tilde U(k+1))=J(x_q(k+1),\tilde U^*(k+1))$. From the algorithm, we have
	\begin{equation}
		\begin{split}
			&\quad J(x_q(k+1),\tilde U^*(k+1))\\
			=&\quad \min\{J(x_q(k+1),\tilde U(k+1)),J(x_q(k+1),\hat U(k+1))\}\\
			\le&\quad  J(x_q(k+1),\tilde U(k+1))\\
			\le&\quad J(x_q(k),\tilde U^*(k)).
		\end{split} 
	\end{equation}
	Therefore, the constructed cost function is a qualified Lyapunov function, and the algorithm converges through iterations.
\end{pf}

\section{Simulation and Analysis}
\label{simulation}
\begin{figure}[h]
	\begin{center}
		\includegraphics[scale=0.285]{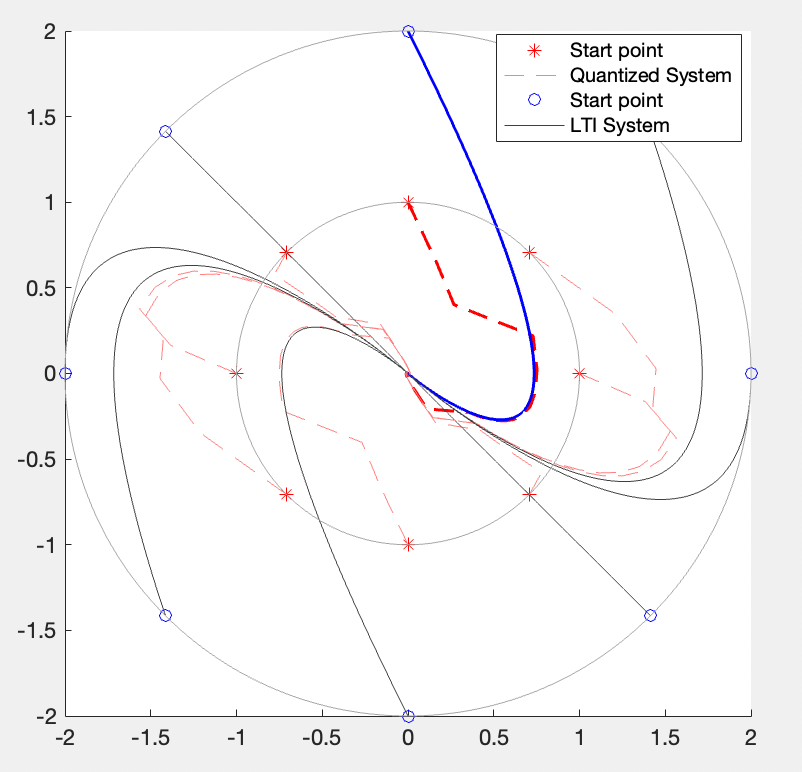}
		\includegraphics[scale=0.25]{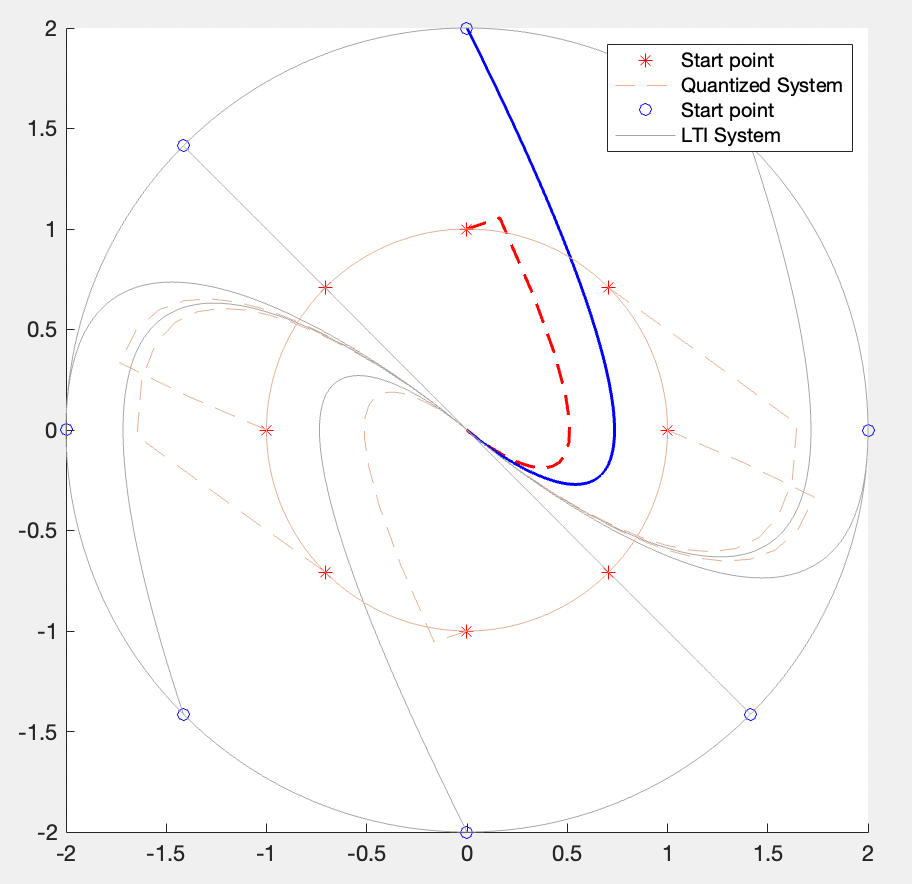}
	\end{center}
	\caption{(a) is the emulation results when solving the integer optimization problem ($\ref{MPC_optimal}$) with the sphere decoding algorithm. (b) applies the suboptimal algorithm $\ref{Suboptimal QS-MPC}$ of the quantized system to emulate a stable LTI system. }
	\label{fig:thm1}
\end{figure}

\begin{figure}[h]
	\begin{center}
		\includegraphics[scale=0.45]{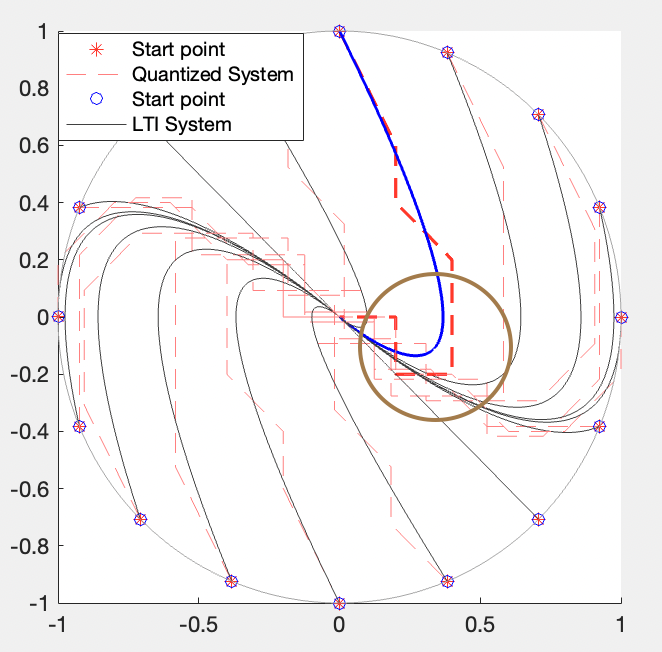}
	\end{center}
	\caption{The emulation result when A is an identity matrix.}
	\label{fig:lemma}
\end{figure}

In this section, we provide simulations of the emulations per Theorem $\ref{MPC:thm1}$ (Fig. \ref{fig:thm1}(a)) and per the Suboptimal QS-MPC of Algorithm $\ref{Suboptimal QS-MPC}$ (Fig. \ref{fig:thm1}(b)) with the same initial conditions. The LTI system we try to emulate is $\dot x=Hx=\begin{bmatrix}
	0& 1\\
	-1&-2
\end{bmatrix}x$ with all its eigenvalues located in the left-half plane. The quantized system we choose is $x_q(k+h)=A_qx(k)+hB_qu(k)=e^{Hh}x_q(k)+h\begin{bmatrix}
	1& 0 & -1 & 0\\
	0&  1&  0&-1
\end{bmatrix}u(k),$ where $u(k)\in\{-1,1,0\}^m$ and time step $h=0.2$. In the cost function, we design $P=\begin{bmatrix}
50 &0 \\
0&50
\end{bmatrix}$, $Q=\begin{bmatrix}
0.1 &0 \\
0&0.1
\end{bmatrix}$ to guarantee the condition (b) in Theorem $\ref{MPC:thm2}$ and Algorithm $\ref{Suboptimal QS-MPC}$ hold. The choice of $P,Q$ also influence the emulation performance: a large diagonal-valued $P$ matrix implies that we are more concerned with predicting the future at the expense of optimality in the current state. To ensure the tracking performance is satisfactory, we choose a proper predicted window size $N=10$. The weight matrix $R$ in the cost function is set to be $0.05*\mathbb{I}_m$. Since if $R$ is too large, the optimal input tends to be zero resulting in no emulation process; if $R$ is zero, it may have singular solutions. We adopt the solver \cite{cplex} to solve the integer programming optimization problem ($\ref{MPC_optimal}$) and obtain the results shown in Fig. $\ref{fig:thm1}$. Gray and blue trajectories are the LTI system starting from the points $\{(2cos\alpha,2sin\alpha):\alpha=0,\pi/4,\cdots,2\pi\}$, while red ones are the quantized system from $\{(cos(\alpha),sin(\alpha)):,\alpha=0,\pi/4,\cdots,2\pi\}$. It can be observed that both quantized systems asymptotically converge to the linear system, although the relaxed constraints emulating algorithm has a larger cumulated error. Meanwhile, we also find the cost function $J$ is strictly decreasing and converging rapidly to zero, which convincingly validates theorem $\ref{MPC:thm2}$. Fig. $\ref{fig:lemma}$ is an emulation with $A=\mathbb{I}_n$ and all initial points on the unit circle. It illustrates that the emulation trajectory is stable but not asymptotically converging to the equilibrium, which is exhibited in the brown circle.

Meanwhile, we have collected data from the left top emulation in Fig. $\ref{fig:thm1}$. There is a total of 3300 data points with quantized directions as their labels. Then we constructed a  regular densely-connected four-layer neural network with ReLU, Sigmoid, or Linear as their activation functions, and the number of nodes in each layer is  512, 480, 256, and 25, respectively. The loss value is calculated by the sparse categorical cross-entropy, which is commonly used in multi-class classification problems, and the optimizer is Adam. After 20 training epochs, we obtain a model with a training accuracy of 98.1\%. The test dataset comes from another emulation with all initial points in the unit circle, which contains 840 data points, and the accuracy achieved 94.7\%. Though using the neural network model to compute the direction has a degraded emulation performance, the running time for training and predicting is 5 minutes, much less than solving integer programming directly, which takes around an hour on the same computer.

%Instead of constructing neural networks, we use logistic regression with a polynomial feature map constructed by the distance and direction for a simple simulation. After training, we obtain a model with a training accuracy of 92\%. The test dataset comes from another emulation with all initial points in the unit circle, which contains 1340 data points, and the accuracy achieved 82\%. Though the regression method has a degraded emulation performance, the running time for training and predicting is 15 minutes, much less than solving integer programming directly, which takes around an hour on the same computer. Because deep neural networks can accurately describe more complexnonlinear relationships, we expect they will provide improved performance. This ongoing research will be described elsewhere.

\section{Conclusion and future work}
The work described above uses concepts of model predictive control (MPC) to extend out
previous research on neuromimetic emulation of finite dimensional linear systems (See \cite{baillieul2021neuromimetic} and \cite{sun2022neuromimetic}). The main results are shown to hold under assumptions that we expect to relax in the near future. Next research will also describe
neuro-inspired machine learning approaches to these and other classes of neuromimetic
emulation.

\begin{ack}
	This work has benefitted from conversations with Anni Li.
\end{ack}
\bibliography{ifacconf}            

\begin{thebibliography}{13}
\providecommand{\natexlab}[1]{#1}
\providecommand{\url}[1]{\texttt{#1}}
\providecommand{\urlprefix}{URL }
\expandafter\ifx\csname urlstyle\endcsname\relax
  \providecommand{\doi}[1]{doi:\discretionary{}{}{}#1}\else
  \providecommand{\doi}{doi:\discretionary{}{}{}\begingroup
  \urlstyle{rm}\Url}\fi

\bibitem[{Baillieul(2019)}]{baillieul2019perceptual}
Baillieul, J. (2019).
\newblock Perceptual control with large feature and actuator networks.
\newblock In \emph{2019 IEEE 58th Conference on Decision and Control (CDC)},
  3819--3826. IEEE.

\bibitem[{Baillieul and Sun(2021)}]{baillieul2021neuromimetic}
Baillieul, J. and Sun, Z. (2021).
\newblock Neuromimetic control—a linear model paradigm.
\newblock In \emph{2021 60th IEEE Conference on Decision and Control (CDC)},
  2709--2716. IEEE.

\bibitem[{Berberich et~al.(2022)Berberich, K{\"o}hler, M{\"u}ller, and
  Allg{\"o}wer}]{berberich2022linear}
Berberich, J., K{\"o}hler, J., M{\"u}ller, M.A., and Allg{\"o}wer, F. (2022).
\newblock Linear tracking mpc for nonlinear systems—part i: The model-based
  case.
\newblock \emph{IEEE Transactions on Automatic Control}, 67(9), 4390--4405.

\bibitem[{Cplex(2013)}]{cplex}
Cplex, I.I. (2013).
\newblock V12. 1: User’s manual for cplex.
\newblock \emph{International Business Machines Corporation}, 46(53), 157.

\bibitem[{Fincke and Pohst(1985)}]{fincke1985improved}
Fincke, U. and Pohst, M. (1985).
\newblock Improved methods for calculating vectors of short length in a
  lattice, including a complexity analysis.
\newblock \emph{Mathematics of computation}, 44(170), 463--471.

\bibitem[{Geyer and Quevedo(2014)}]{geyer2014multistep}
Geyer, T. and Quevedo, D.E. (2014).
\newblock Multistep finite control set model predictive control for power
  electronics.
\newblock \emph{IEEE Transactions on power electronics}, 29(12), 6836--6846.

\bibitem[{Gr{\"o}tschel et~al.(2012)Gr{\"o}tschel, Lov{\'a}sz, and
  Schrijver}]{grotschel2012geometric}
Gr{\"o}tschel, M., Lov{\'a}sz, L., and Schrijver, A. (2012).
\newblock \emph{Geometric algorithms and combinatorial optimization}, volume~2.
\newblock Springer Science \& Business Media.

\bibitem[{Hassibi and Vikalo(2005)}]{hassibi2005sphere}
Hassibi, B. and Vikalo, H. (2005).
\newblock On the sphere-decoding algorithm i. expected complexity.
\newblock \emph{IEEE transactions on signal processing}, 53(8), 2806--2818.

\bibitem[{Karamanakos et~al.(2015)Karamanakos, Geyer, and
  Kennel}]{karamanakos2015computationally}
Karamanakos, P., Geyer, T., and Kennel, R. (2015).
\newblock A computationally efficient model predictive control strategy for
  linear systems with integer inputs.
\newblock \emph{IEEE Transactions on Control Systems Technology}, 24(4),
  1463--1471.

\bibitem[{Lewis and Tou(1963)}]{lewis1963optimum}
Lewis, J. and Tou, J. (1963).
\newblock Optimum sampled-data systems with quantized control signals.
\newblock \emph{IEEE Transactions on Applications and Industry}, 82(67),
  229--233.

\bibitem[{Liu and Skelton(1990)}]{liu1990optimal}
Liu, K. and Skelton, R. (1990).
\newblock Optimal controllers for finite wordlength implementation.
\newblock In \emph{1990 American Control Conference}, 1935--1940. IEEE.

\bibitem[{Sun and Baillieul(2022)}]{sun2022neuromimetic}
Sun, Z. and Baillieul, J. (2022).
\newblock Neuromimetic linear systems—resilience and learning.
\newblock In \emph{2022 IEEE 61st Conference on Decision and Control (CDC)},
  7388--7394. IEEE.

\bibitem[{Xu et~al.(2022)Xu, Damsma, and Lazar}]{xu2022steady}
Xu, D., Damsma, S., and Lazar, M. (2022).
\newblock On the steady-state behavior of finite-control-set mpc with an
  application to high-precision power amplifiers.
\newblock In \emph{2022 European Control Conference (ECC)}, 820--825.
\newblock \doi{10.23919/ECC55457.2022.9838191}.

\end{thebibliography}

\appendix
\section{Proof of Theorem \ref{MPC:thm1}} 
\begin{figure}[h]
	\begin{center}
		\includegraphics[scale=0.4]{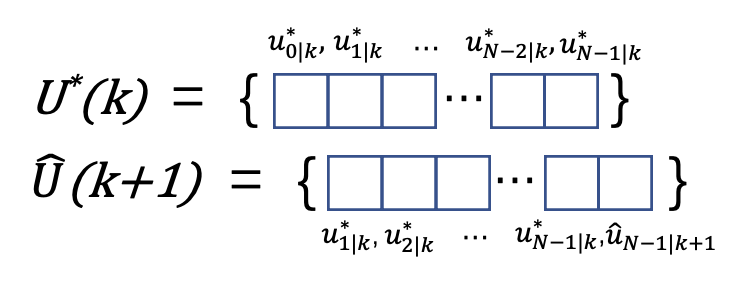}
	\end{center}
	\caption{The shifted input sequence $\hat U(k+1)$ and $U^*(k)$. }
	\label{fig:shift}
\end{figure}
	Assume the optimal input sequence $U^*(k)=\{u^*_{0|k},...,$ $u^*_{N-1|k}\}$ solves the optimization problem (\ref{MPC_optimal}) at time $k$ and apply $u^*_{0|k}$ to the quantized system (\ref{equation:qs}). At time $k+1$, we define a shifted input sequence $\hat U$ to be 
	\begin{equation}
		\begin{split}
			\hat U(k+1)&\quad =\quad \{\hat u_{0|k+1},...,\hat u_{N-2|k+1},\hat u_{N-1|k+1}\}\\
			&\quad =\quad \{u^*_{1|k},...,u^*_{N-1|k},\hat u_{N-1|k+1}\},
		\end{split}
		\label{shifted U}
	\end{equation}
	where $\hat u_{N-1|k+1}\in \{-1,0,1\}^m$ is the predicted next $N-1$ input  at time $k+1$, which has not been defined up to this point. Figure($\ref{fig:shift}$) exhibits the relationship between $\hat U(k+1)$ and $U^*(k)$.
	Then, $J(x_q(k+1),\hat U(k+1))$ can be expressed as 
	\begin{equation}
		\begin{split}
			&J(x_q(k+1),\hat U(k+1))=J(x_q(k),U^*(k))-|u^*_{0|k}|^2_R\\
			&\quad\quad\quad\quad\quad-|x_{0|k}-x_{ref}(0,k)|^2_Q-|x_{N|k} -x_{ref}(N,k)|^2_P\\
			&\quad\quad\quad\quad\quad+|x_{N-1|k+1}-x_{ref}(N-1,k+1)|^2_Q\\
			&\quad\quad\quad\quad\quad+|x_{N|k+1} -x_{ref}(N,k+1)|^2_P+|\hat u_{N-1|k+1}|^2_R.
		\end{split}
		\label{shifted J}
	\end{equation}
	 Since the second term ($-|u^*_{0|k}|^2_R$) and third term ($-|x_{0|k}-x_{ref}(k,0)|^2_Q$) in the above polynomial are always smaller or equal than zero, the sum of the last four terms in (\ref{shifted J}) is non–positive providing a sufficient condition to guarantee the optimal cost function $J$ is non-increasing at time $k + 1$. Meanwhile, it should be noted that the optimal input sequence $U^*(k+1)=\{ u^*_{0|k+1},...,u^*_{N-2|k+1}, u^*_{N-1|k+1}\}$ obtained by solving the optimization problem (\ref{MPC_optimal}) at time $k+1$ should have an equal or lower cost than any other quantized inputs including $\hat U(k+1)$, i.e.,
	\begin{equation}
		\label{optimal J}
		\begin{split}
			J(x_q(k+1),U^*(k+1))\le J(x_q(k+1),\hat U(k+1)).
		\end{split}
	\end{equation}
	Therefore, from (\ref{shifted J}) and (\ref{optimal J}), we can obtain
	\begin{equation}
		\label{decrease J}
		\begin{split}
			&J(x_q(k+1),U^*(k+1))-J(x_q(k),U^*(k))\\ 
			\le &J(x_q(k+1),\hat U(k+1))-J(x_q(k),U^*(k))\\
			\le &-|x_{0|k}-x_{ref}(k,0)|^2_Q-|u^*_{0|k}|^2_R\le0,
		\end{split}
	\end{equation}
	as long as the sum of the last four terms in (\ref{shifted J}) is non–positive.
	
	In addition, since $J(x_q(k+1),U^*(k+1))-J(x_q(k),U^*(k))$ is less or equal to two non-positive terms exhibited in (\ref{decrease J}), $J(x_q(k+1),U^*(k+1))-J(x_q(k),U^*(k))$ equals 0 if and only if both of the two terms are zero, i.e., $u^*_{0|k}=\bold{0}$, $x_{0|k}=x_{ref}(0,k)$ in (\ref{decrease J}). When the optimal $u^*_{0|k}$ is applied to the quantized system at time $k$ and assuming $A_q=e^{Hh}$, $x_q(k+1)=A_qx_q(k)=A_qx_{0|k}=e^{Hh}x_{0|k}$ for the quantized system with zero input. At the same time, the LTI states become $x_{ref}(0,k+1)=e^{Hh}x_{ref}(0,k)=e^{Hh}x_{0|k}=x_q(k+1)$. Because the quantized system can always track the LTI system without inputs going forward from time $k$, the optimal input sequence will be zero when solving the optimization problem (\ref{MPC_optimal}). By observing the structure of the non-negative function $J$, we can conclude that it will be zero from time $k$ onward. Therefore, the cost function is strictly decreasing until it becomes zero.
	%It is observed that $J(x(k),U^*(k))$ is strictly decreasing until it becomes 0, i.e., if $J(x(k+1),U^*(k+1))-J(x(k),U^*(k))=0$, then for all $l\ge0,$ $J(x(k+l),U^*(k+l))=0$. It is because $J(x(k+1),U^*(k+1))-J(x(k),U^*(k))$ is less or equal than two non-positive terms. $J(x(k+1),U^*(k+1))-J(x(k),U^*(k))=0$ if and only if the two non-positive terms in (\ref{decrease J}) $-|x_{0|k}-x_{ref}(k,0)|^2_Q$ and $-|u^*_{0|k}|^2_R$ are both zero. Then, $u^*_{0|k}=\bold{0}$, $x_{0|k}=x_{ref}(k,0)$ and $u^*_{0|k}$ would apply to the quantized system at time $k$. For the quantized system, $x(k+1)=Ax(k)=Ax_{0|k}$, and the LTI states become $x_{ref}(k+1,0)=e^Hx_{ref}(k,0)=e^Hx_{0|k}=x(k+1)$. When solving the optimization problem (\ref{MPC_optimal}), the optimal input sequence will be zero because the quantized system can always tracking the LTI system without inputs from time $k$. By observing the structure of non-negative function $J$, we can find it will also be zero from time $k$. Therefore, the cost function is strictly decreasing until it becomes zero, so that it satisfies the definition of a qualified Lyapunov function.\\
	
	Next, we will prove the conditions stated in the Theorem \ref{MPC:thm1} are sufficient to ensure the sum of the last four terms in (\ref{shifted J}) are non–positive. Since $\hat u_{N-1|k+1}$ is not initially defined in $\hat U(k+1)$, we can assign it to be $\bold{0}$. Meanwhile, $x_{ref}(N,k+1)=e^{Hh}x_{ref}(N,k)$, $x_{ref}(N,k)=x_{ref}(N-1,k+1)$, $x_{N-1|k+1}=x_{N|k}$ and $x_{N|k+1}=A_qx_{N-1|k+1}+B_q\hat u_{N-1|k+1}=A_qx_{N-1|k+1}=A_qx_{N|k}$. Then, the last four terms in (\ref{shifted J}) can be written as 
	\begin{equation}
		\begin{split}
			&|x_{N-1|k+1}-x_{ref}(N-1,k+1)|^2_Q-|x_{N|k} -x_{ref}(N,k)|^2_P\\
			&\qquad\qquad\quad+|x_{N|k+1} -x_{ref}(N,k+1)|^2_P+|\hat u_{N-1|k+1}|^2_R\\
			&=|A_qx_{N|k} -e^{Hh}x_{ref}(N,k)|^2_P-|x_{N|k} -x_{ref}(N,k)|^2_{P-Q}
		\end{split}
	\end{equation}
	If $A=e^{Hh}$, the above equation can be written as\[(x_{N|k} -x_{ref}(N,k))^T(A_q^TPA_q+Q-P)(x_{N|k} -x_{ref}(N,k)).\] 
	When $Q-P+A_q^TPA_q\prec0$ with $P,Q$ are positive definite, for any prediction horizon $N$, the system (\ref{equation:qs}) is asymptotically stable.
	
	%%%%%%%%%%%%%%%%%%%%%%%%%%%%%%%%%%%%%%%%%%%%%%%%%
	\section{Transform the MPC objective to integer Least-square form}
	Following the standard construction of MPC, we write our system in the extensive form:
	\begin{equation}
		\begin{split}
			x_{0|k}& = x_q(k),\\ 
			x_{1|k}& = A_qx_q(k)+B_qu(k),\\
			& \vdots\\
			x_{N|k}& = A_q^Nx_q(k)+\sum_{i=0}^{N-1}A_q^{N-1-i}B_qu(k+i|k).
		\end{split}
	\end{equation}
	Therefore, QS-MPC equations can be written as $X_k=\tilde{A}x_q(k)+\tilde{B}U_k$ and the objective function in $(\ref{MPC_optimal})$ can be rewritten as
	\begin{equation} 
		\begin{split}
			J&=(X_k-R_k)^T\tilde Q(X_k-R_k)+U_k^T\tilde RU_k\\
			& =(\tilde Ax_q(k)+\tilde BU_k-R_k)^T\tilde Q(\tilde Ax_q(k)+\tilde BU_k-R_k)\\
			&\ \ \ \ \ \ \ \ \ \ \ \ \ \ \ \ \ \ \ \ \ \ \ \ \ \ \ \ \ \ \ \ \ \ \ \ \ \ \ \ \ \ \ \ \ \ \  \ \ \ \ \ \ \ \ \ \ +U_k^T\tilde RU_k\\
			& = (\tilde Ax_q(k)-R_k)^T\tilde Q(\tilde Ax_q(k)-R_k)\\
			&\ \ \ \ \ \ \ \ \ \ \ \ \ \ \ \ \ \ \ \ \ \ +2(\tilde Ax_q(k)-R_k)^T\tilde Q\tilde BU_k+U_k^T\tilde HU_k\\
			& =U_k^T\tilde HU_k + 2(\tilde Ax_q(k)-R_k)^T\tilde Q\tilde BU_k + Const.(k)\\
			& = (U_k-U_{uncon K})^T\tilde H(U_k-U_{uncon K})+ Const.'(k),
		\end{split}
	\end{equation} 

where $\tilde H=\tilde B^T\tilde Q\tilde B+\tilde R$, $U_{uncon K}=-\tilde H^{-1}$ $(\tilde Ax_q(k)-R_k)$. Therefore, the problem $(\ref{MPC_optimal})$ can be written as 
	\begin{equation}
		\begin{split}
			\min\limits_{U_k}\quad& ||U_k+\tilde H^{-1}(\tilde Ax_q(k)-R_k)||_{\tilde H}^2\\
			s.t. \quad &u_{n|k}\in\left\{-1,\ 0,\ 1\right\}^m,\ \ \ \ \ \forall n= {0,1,...,N-1}. 
		\end{split}
	\end{equation}
\end{document}